\documentclass[aps,apl,twocolumn]{revtex4}

\usepackage{amsmath,cases}
\usepackage{amsmath}
\usepackage{amssymb}
\usepackage[dvipdfmx]{graphicx}  
\usepackage[dvipdfmx]{color}
\usepackage{verbatim}

\newcommand{\bra}[1]{\langle{#1}|}
\newcommand{\ket}[1]{|{#1}\rangle}

\begin{document}

\title{Reachable Set Characterization of Open Quantum System by Quantum Speed Limit}
\author{Kohei Kobayashi}
\affiliation{Quantum Computing Center, Keio University, 
Hiyoshi 3-14-1, Kohoku, Yokohama, 223-8522, Japan}
\date{\today}

\begin{abstract}
In recent years, Arenz et al. proposed the idea of reachable set characterization based on the quantum speed limit (QSL); that is,
the reachable set of the target unitary gate in a closed qubit system can be characterized by considering the QSL as the necessary condition that the control setup must satisfy in order to achieve the goal. 
Inspired by this idea, in this paper we characterize a general Markovian open quantum system based on the QSL derived in \cite{Kohei2}. 
Note that this bound is not only explicitly computable with respect to system parameters, but also tighter than the other bounds.  
Some examples for demonstrating this analysis will be given.

\end{abstract}

\maketitle


\section{Introduction}

Quantum information technologies, e.g., quantum computing, communication, teleportation, and metrology, is receiving a lot of attention in recent years, due to its possibility to outperform classical information processing.
These technologies are described by a sequence of qubits for an information resource.
Therefore a technique of quantum control which efficiently and accurately prepare the desired quantum states plays an important role. 
However, in the practical situation, there is an undesirable environment effect such as {\it decoherence} which destroys the coherence properties of quantum systems, 
resulting that the actual control performance severely degrades.
Thus, it is important to quantitatively evaluate and characterize a reachable set of the controlled state, 
i.e, the set of all the final states  reachable starting from the initial state, from the perspective of quantum engineering.  
In general, the reachable set is described as a time-dependent function, e.g., Lie-semigroup structures \cite{Dirr}, C-numerical range \cite{Dirr2},  
spectrum of the density matrix \cite{Yuan}, and so on. 
In order to study this problem, the {\it quantum speed limit} (QSL) can be used as a useful tool.  \\
\ \ \ The QSL is defined as a fundamental lower bound on the evolution time of a quantum state from an initial state to a final state.  
The study of the QSL started from closed systems; Mandelstam and Tamm presented a first QSL between orthogonal states \cite{MT}, 
which is bounded by the variance of the system energy.
Later decades, Margolus and Levitin provided another QSL depending on mean energy \cite{ML}, and 
extensions to mixed states \cite{mixed}, nonorthogonal states \cite{nonorthogonal1, nonorthogonal2}, 
time-dependent driven system \cite{timedependent1, timedependent2}, and 
open quantum systems \cite{Taddei, Campo, Deffner, Sun, Meng, Zhang} have been investigated. 
The QSL gives a trade-off relation between the control time and energy resource, 
and thus it has several applications in quantum control scenarios \cite{Caneva, Poggi1, Andersson, Poggi2, Poggi3}.
Actually, the connection between the reachable set and the QSL has been studied in recent years.
Arenz et al., characterized the sets of the all target unitary gates by considering the QSL for the control time as the necessary condition that control parameters, e.g., the drift and control Hamiltonians and driving time, must satisfy in order to implement the target gate \cite{Arenz}.  \\
\ \ \ Inspired by the above facts, in this paper we extend the work by Arenz et al. to general Markovian open quantum systems.
More precisely, we characterize the reachable set of an open quantum system under a certain situation, e.g., initial state, Hamiltonian, decoherence, and evolution time, using the QSL.
In order to make this analysis tractable and rigorous, we employ the bound derived in \cite{Kohei2}, because it is not only explicitly computable with respect to control parameters, but also tighter than the other explicit bounds. 
Sec. III provides some examples for demonstrating this analysis with insightful notions.

\section{Setup and QSL}

In this section, we introduce the QSL presented in \cite{Kohei2}. 
We first begin with considering the Markovian master equation
\begin{align}
\label{ME}
\frac{d\rho_t }{dt}=-i[H, \rho_t]+ {\mathcal D}[M]\rho_t,
\end{align}
where $H$ is the time-independent Hamiltonian. 
$M$ is the Lindblad operator representing the decoherence, and thus ${\mathcal D}[M]\rho=M\rho M^\dagger-M^\dagger M\rho/2-\rho M^\dagger M/2$.
Throughout this paper, we assume that $\hbar=1$ and the initial state is pure, i.e., $\rho_0=\ket{\psi_0}\bra{\psi_0}$.
Next we focus on the relative purity $\Theta_t$ between initial state and final state
\begin{align}
\label{relative}
\Theta_t:={\rm arccos}\left( \langle\psi_0|\rho_t |\psi_0\rangle \right), 
\end{align}
where $0\leq \Theta_t \leq \pi/2$.
The relative purity has been often employed for deriving the QSL \cite{Campo, Meng}.  
The upper bound is achieved when $\rho_0$ and $\rho_t$ are orthogonal, and the lower one is achieved only when $\rho_t=\rho_0$.
In this setting, we find that the dynamics of $\Theta_t$ is upper bounded as follows:
\begin{align}
\label{dynamics}
\frac{d\Theta_t}{dt} \leq \frac{1}{\sin\Theta_t}(\mathcal{A}\sqrt{1-\cos\Theta_t }+\mathcal{E}),
\end{align}
with
\begin{align*}
     &\mathcal{A}=\sqrt{2}\left( \|  i[H, \rho_{0}]+\mathcal{D}^{\dagger}[M]\rho_0 \|_{\rm F}\right),  \\   
     &\mathcal{E}=\|M\ket{\psi_{0}}\|^{2}-|\bra{\psi_{0}}M\ket{\psi_{0}}|^{2},   
\end{align*}
where  $\|X\|_{\rm F}=\sqrt{{\rm Tr}(X^\dagger X)}$, $\|\ket{\psi}\|=\sqrt{ \langle\psi|\psi \rangle}$, and
$\mathcal{D}^{\dagger}[M]\rho=M^\dagger \rho M-M^\dagger M\rho/2-\rho M^\dagger M/2$.
Integrating the above inequality \eqref{dynamics} from $0$ to $T$, we obtain the following  lower bound \cite{Kohei2}:
\begin{align}
\label{QSL}
T\geq T_*:=\frac{2 \lambda }{\mathcal{A}}+\frac{2\mathcal{E}}{\mathcal{A}^2}\ln\left(\frac{\mathcal{E} }{\mathcal{E}+\mathcal{A} \lambda }  \right),
\end{align}
with $\lambda=\sqrt{1- \cos\Theta_T }$ $(0 \leq \lambda \leq 1)$.
The derivation of the above results are given in Appendix A.
This upper bound is achieved when $\Theta_T=\pi/2$ and the lower one is achieved when $\Theta_0=0$, 
Hence $\lambda$ can be interpreted as the radius of the ball around $\rho_0$.
The QSL $T_*$ gives a lower bound on the time $T$ for the state $\rho_t$ to evolve from $\rho_0$ to $\rho_T$ (i.e., $\Theta_t$ evolves from $0$ to $\Theta_T$). 
As shown in Ref \cite{Kohei2}, $T_*$ has the following notable features: 
(i) $T_*$ is applicable to a general Markovian open quantum system driven by arbitrary $H$ and $M$.
(ii) $T_*$ is explicitly computable once the parameters $(\rho_0, H, M, \Theta_T)$ are specified. 
Thanks to this, it is not necessary for solving any equations to calculate $T_*$. 
(iii) When $M=M^\dagger$, $H=0$, and $\ket{\psi_0}$ is an eigenstate of $M$, or   
$M=0$ and $[H, \rho_0]=0$, we have $T_*\to \infty$ due to $\mathcal{A}=\mathcal{E}=0$.
In this case, the state cannot reach any $\rho_T\neq \rho_0$, resulting that $\lambda$ remains zero.  
(iv) $T_*$ is tighter than the other explicit bound;
for the same setup mentioned above, del Campo provided the QSL: $T\geq T_{\rm DC}:=\sqrt{2}\lambda/\mathcal{A}$ \cite{Campo}.
Note that $T_* \geq T_{\rm DC}$ holds when the decoherence is small or the radius $\lambda$ is small. 
(v) If $H$ is time-dependent, we cannot generally solve the integral \eqref{dynamics}. 
Then, defining $H_t=u_tH'$ where $u_t$ is the time-dependent control input with a energy constraint $u_{\rm max}:={\rm max}\{|u_t|\}$ and using the triangle inequality,
${\mathcal A}$ is redefined as follows:
\begin{align}
\mathcal{A}'=\sqrt{2}\left( u_{\rm max}\|  i[H', \rho_{0}]\|_{\rm F} + \|\mathcal{D}^{\dagger}[M]\rho_0 \|_{\rm F}\right).
\end{align}
Therefore, in this case the QSL is given by \eqref{QSL} replacing $\mathcal{A}$ with $\mathcal{A}'$.   \\
\ \ \ In addition to the above points, here we explain the idea of charcterization of the reachable set of targets based on the QSL;
the parameters determining the state evolution $(\rho_0, H, M, T)$ and the fidelity-based distance $\lambda$ are correlated via the inequality $T\geq T_*$. 
In other words, this relation can be considered as a necessary condition that $(\rho_0, H, M, T)$ must satisfy in order to achieve a certain $\lambda$.
Therefore, once these parameters are specified, we can characterize the set of all $\lambda$ (i.e., the set of all final states $\ket{\psi_T}$) by examining $T\geq T_*$.
Now as a simple example, let us consider the closed system driven by $H$. 
In this case, we have $\mathcal{E}=0$ and 
\begin{align*}
\mathcal{A} &=\sqrt{2}\| i[H,\rho_0]\|_{\rm F} =2\sqrt{ \langle \psi_0|H^2|\psi_0\rangle-\langle \psi_0|H|\psi_0\rangle^2 },
\end{align*}
which corresponds to the variance of the system energy in the initial state.
Hence for the state evolution from $\ket{\psi_0}$ to any final state $\ket{\psi_T}$, $\lambda=\sqrt{1-\cos\Theta_T}$ 
has an upper bound 
\begin{align}
\label{upperbound}
\lambda \leq \sqrt{ \langle \psi_0|H^2|\psi_0\rangle-\langle \psi_0|H|\psi_0\rangle^2 } T.
\end{align} 
In particular, when $T$ is small such that the rightmost side of \eqref{upperbound} is less than one, 
it gives a meaningful bound with respect to $\lambda$.
In what follows, some typical examples demonstrating this idea will be given.

\section{Example}

\subsection{Qubit}

First we study a qubit system consisting of the excited state $\ket{0}=[1, 0]^\top$ and  the ground state $\ket{1}=[0, 1]^\top$.
Its initial pure state can be fully parametrized by the Bloch representation:
\begin{align}
\label{initial}
\ket{\psi_0}=[\cos\theta,\ e^{i\varphi}\sin\theta]^\top, 
\end{align}
where $0\leq \theta, \varphi \leq \pi$.
Also the system observables along the $i$ axis ($i=x, y, z$) are given by the Pauli matrices
$\sigma_x=\ket{0}\bra{1}+\ket{1}\bra{0}$, $\sigma_y=-i\ket{0}\bra{1}+i\ket{1}\bra{0}$,
and $\sigma_z=\ket{0}\bra{0}-\ket{1}\bra{1}$.
Here we consider the following setup:
\begin{align}
\label{setup}
H= \Omega \sigma_z, \ M=\sqrt{\gamma}\sigma_-.
\end{align}
$H$ represents the rotation the state along the $z$ axis with the frequency $\Omega>0$ and 
$M$ represents the energy decay $\ket{0}\to\ket{1}$ with decay rate $\gamma>0$. 
In this case, the lower bound is given by \eqref{QSL} with

\begin{align*}
\mathcal{A} &= \sqrt{2\gamma^2\cos^22\theta+(4\Omega^2+\frac{\gamma^2}{4})\sin^22\theta },  \\
\mathcal{E} &=\gamma\cos^4\theta.
\end{align*}

First we focus on the case of $\gamma=0$, where $\lambda$ has an upper bound 
$\lambda \leq \Omega|\sin2\theta|T$.
Figure 1 (a) shows the reachable set of $\lambda$ for each initial states and driving times.
The colored area is the all sets satisfying this inequality and the white area is the one that does not satisfy it.
We find that the three curves takes zero at $\theta=0$ and $\pi/2$, meaning that $\ket{0}$ and $\ket{1}$ does not change under the rotation.
This is consistent with the fact that $\ket{0}$ and $\ket{1}$  are the steady state of 
the equation $d\rho_t/dt=-i\Omega[\sigma_z, \rho_t]$.
Further, the reachable sets take maximum at the superposition $\ket{+}:=(\ket{0}+\ket{1})/\sqrt{2}$, which lies on the equator of the Bloch sphere.
This implies that the rate of the state change from $\ket{+}$ is biggest, and thus is the most fragile state against the rotation induced by $H$.
On the other hand, when $\gamma=1$, the curves take the maximum at $\theta=0$ [Fig. 1 (b)].
This is because $\ket{0}$ is largely affected by the decoherence $M$, while it is unchanged against $H$ due to the fact that $\ket{0}$ is the eigenstate of $H$.
In particular, it is noted that the state cannot reach its orthogonal state within $T=0.1$, $0.3$, 
while $\ket{\psi_0}$ with $\theta \leq 0.9$ may do so within $T=0.8$. \\
\ \ \ Next we consider an unitary gate implementation; that is,
we aim to implement the target gate by  appropriately designing the time-dependent Hamiltonian $H_t$ under the assumption that $M=0$.
Now let the target gate be the rotation operator 
$G(\alpha, \beta, \delta)=R_z(\alpha)R_y(\beta)R_z(\delta)$ with
\begin{align*}
&R_z(\alpha)=e^{-i\frac{\alpha}{2}\sigma_z}= 
\left[
    \begin{array}{cc}
      e^{-i\frac{\alpha}{2} }  & 0 \\
     0  & e^{i\frac{\alpha}{2} }
    \end{array}
  \right],   \\
&R_y(\beta)=e^{-i\frac{\beta}{2}\sigma_y}= \left[
    \begin{array}{cc}
     \cos\frac{\beta}{2}   & -\sin\frac{\beta}{2} \\
     \sin\frac{\beta}{2}  & \cos\frac{\beta}{2}
    \end{array}
  \right],   \\  
&R_z(\delta)=e^{-i\frac{\delta}{2}\sigma_z}
= \left[
    \begin{array}{cc}
      e^{-i\frac{\delta}{2} }  & 0 \\
     0  & e^{i\frac{\delta}{2} }
    \end{array}
  \right],   
\end{align*}
where $0 \leq\alpha <2\pi$, $0 \leq\beta <\pi$, and $0 \leq\delta <4\pi$.
$R_y$ and $R_z$ represent the rotation around $y$ and $z$ axis, respectively, and 
the transformation from $\ket{\psi_0}$ to $\ket{\psi_T}$ is described by $\ket{\psi_T}=G(\alpha, \beta, \delta)\ket{\psi_0}$. 
Also we take the Hamiltonian as
\begin{align}
\label{Hamiltonian}
H_t =\Omega \sigma_x+u_t \sigma_z.
\end{align}
Note that the state is fully controllable, hence every $G(\alpha, \beta)\subset {\rm SU}(2)$ can be implemented by suitably choosing $u_t$  for sufficient time \cite{Arenz}.
In this case the lower bound  is calculated as follows:

\begin{align}
\label{QSLgate}
T_\ast
=\frac{ \sqrt{ 1-\cos^2\frac{\alpha}{2} \cos^2\frac{\beta}{2}-\sin^2\frac{\alpha}{2}\cos^2( 2\theta+\frac{\beta}{2} ) } }{ \Omega|\cos2\theta|+ u_{\rm max}|\sin2\theta|}.
\end{align}
If $\theta=0$, $T_*=|\sin(\beta/2)|/\Omega$.
Figure 1 (c) shows the sets that satisfies $T\geq T_*$
for $T=0.3$, $0.5$, and $0.8$ under the fixed values $\Omega=u_{\rm max}=1$. 
Note that for instance, when $T=0.5$, this inequality is saturated at $\beta=\pi/3$, where the target is given by
\begin{align*}
G(\alpha) = \frac{1}{2}\left[
    \begin{array}{cc}
      \sqrt{3}e^{-i\frac{\alpha}{2} }  & -e^{-i\frac{\alpha}{2} } \\
      e^{i\frac{\alpha}{2}} & \sqrt{3}e^{i\frac{\alpha}{2} }
    \end{array}
  \right].
\end{align*}
This $G$ gives the final fidelity $\cos\Theta_T=0.75$. 
Then we can only steer the state to the one with fidelity about $0.75$ at most in $T=0.5$. 
Next for the case of $\theta=\pi/4$, the lower bound is 
$T_*=|\sin(\alpha/2-\beta/2) |/u_{\rm max}$.
Clearly, the reachable sets depicted in Fig. 1 (d) are remarkably changed from Fig. 1 (c).
The lower bound $T_*(\alpha, \beta)$ takes zero at $G(0, 0)=G(2\pi, 0)=I$ and $G(\pi, \pi)=i\sigma_x$, because $\ket{\psi_0}=\ket{+}$ is the eigenstate of $\sigma_x$. 
This means that there is no difference between $\ket{\psi_0}$ and $\ket{\psi_T}$ in terms of the fidelity.
On the other hand, $T_*(\alpha, \beta)$ is maximized at 
\begin{align}
\label{gate}
G(0, \pi)=G(2\pi, \pi)=-i\sigma_y, \ G(\pi, 0)=i\sigma_z.
\end{align}
Actually, they cannot be implemented even if $T=0.8$.
Indeed, this result is reasonable because the target states generated by $-i\sigma_y$ and $i\sigma_z$ are given as
$[1, -1]/\sqrt{2}$ and $[i, -i]/\sqrt{2}$, which are orthogonal to $\ket{\psi_0}$. 
Therefore, we can conclude that it takes longer time to prepare the state that is far from $\ket{\psi_0}$.
As shown above, by using $T_*$, we can characterize the set of the target states or target gates under a given control setting.

\begin{figure}[tb]
\centering
\includegraphics[width=8.5cm]{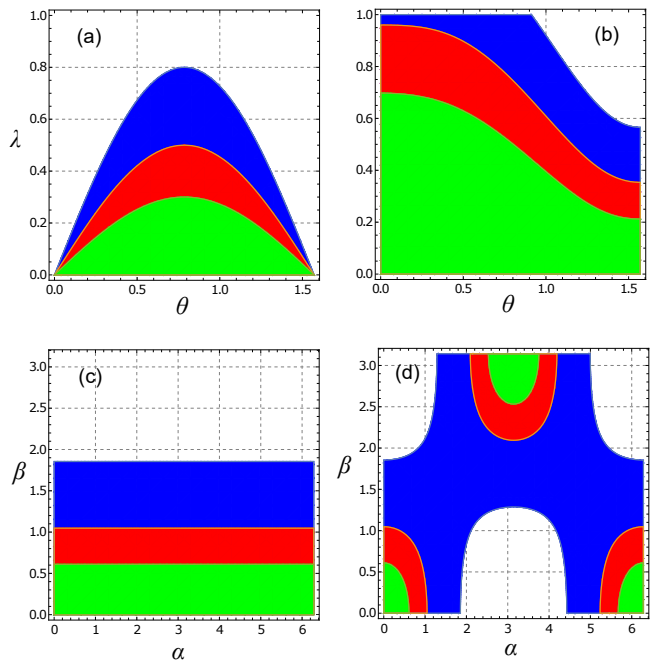}
\caption{Reachable set of $\lambda$  as a function of $\theta$ for each evolution times $T=0.3$ (green area), $T=0.5$ (red area), and $T=0.8$ (blue area)
 when (a) $\gamma=0$ and (b) $\gamma=1$, in unit of $\Omega=1$.
Reachable set of the target gate $G(\alpha, \beta)$ for $T=0.3$ (green area), $T=0.5$ (red area), and $T=0.8$ (blue area)  when (a) $\theta=0$ and (b) $\theta=\pi/4$, in unit of $\Omega=u_{\rm max}=1$.}
\end{figure}

\subsection{Two qubits}
Here we study a two-qubit system driven by decoherence.
We first consider the Bell states, which is the maximally entangled states defined as
\begin{align*}
&\ket{\Phi^\pm}=\frac{1}{\sqrt{2}}(\ket{0}\otimes\ket{0}+\ket{1}\otimes\ket{1}),\\
& \ket{\Psi^\pm}=\frac{1}{\sqrt{2}}( \ket{1}\otimes\ket{0}+\ket{0}\otimes\ket{1}).
\end{align*}
The Bell states are used as a main resource for the quantum information processing such as quantum teleportation.
Thereby, it is of particular interest in comparing the sets of the state from each Bell states under undesirable noises.
Now we take the collective (global) noise modeled by $M=\sqrt{\gamma}(\sigma_-\otimes I+I\otimes \sigma_-)$.
Further, for simplicity, we assume $H=0$. Then, the lower bound for each states are calculated as follows:
\begin{align*}
 &T_*(\ket{\Phi^\pm})=\frac{2\lambda}{\sqrt{5}\gamma}-\frac{2}{5\gamma}\ln(1+\lambda),  \\
 &T_*(\ket{\Psi^+})=\frac{\lambda}{2\gamma}-\frac{1}{4\gamma}\ln(1+2\lambda),  
\end{align*}
and $T_*(\ket{\Psi^-})\to\infty$ due to $\mathcal{A}=0$. 
This corresponds to the fact that the initial state $\ket{\Psi^-}$ does not change
because it is identical to the zero eigenstate of $M$.
Hence, as an information resource, $\ket{\Psi^-}$ is the best initial state  against $M$.
Also, as shown in Fig. 2 (a), the set of the state from $\ket{\Psi^+}$ spreads faster than that of $\ket{\Phi^\pm}$ as $\gamma$ increases, implying that $\ket{\Phi^\pm}$ can preserve its coherence longer than $\ket{\Psi^+}$.      \\
\ \ \ Next we consider a qutrit, which is a symmetric two-qubit system described by a three-level system
consisting of the distinguishable states $\ket{E}=[1, 0, 0]^\top$, $\ket{S}=[0, 1, 0]^\top$, and $\ket{G}=[0, 0, 1]^\top$.
Note that $\ket{S}$ corresponds to the entangled state between two states.
As seen in the qubit example, we investigate a reachable set of the target $3\times 3$ gate.
Let us consider a rotation operator of the three-dimensional system $G(\alpha, \beta, \delta)=R_z(\delta)R_y(\beta)R_x(\alpha)$
where matrix representations of $R_i$ ($i=x, y, z$) are given as
\begin{align*}
&R_x(\alpha) = \left[
    \begin{array}{ccc}
       \cos\alpha & -\sin\alpha & 0 \\
      \sin\alpha &  \cos\alpha  &  0 \\
      0 &  0 &  1
    \end{array}
  \right], \\ 
 &  R_y(\beta) = \left[
    \begin{array}{ccc}
      \cos\beta &  0&\sin\beta  \\
      0 &   1 &  0 \\
      -\sin\beta & 0  &  \cos\beta
    \end{array}
  \right], \\   
&  R_z(\delta) = \left[
    \begin{array}{ccc}
    1 &0 &0  \\
    0&   \cos\delta & -\sin\delta  \\
    0&  \sin\delta &  \cos\delta 
    \end{array}
  \right], 
\end{align*}
where $\alpha$, $\beta$, and $\delta$ are the rotation angles about $x$, $y$, and $z$-axis.
We limit the initial pure state to the real vector
\begin{align*}
\ket{\psi_0}=[ \sin(\theta/2)\cos(\varphi/2),\ \cos(\theta/2), \ \sin(\theta/2)\sin(\varphi/2)]^\top,
\end{align*}
where $0\leq \theta$, $\varphi\leq \pi$. 
Also we assign the following Hamiltonian:
\begin{align}
\label{qutritHam}
H_t = \Omega S_x + u_tS_z,
\end{align}
where $S_x=(\ket{S}\bra{E}+\ket{G}\bra{S})/\sqrt{2}$+H.c., 
$S_y=i(\ket{S}\bra{E}+\ket{G}\bra{S})/\sqrt{2}$+H.c., and $S_z=\ket{E}\bra{E}-\ket{G}\bra{G}$
are the spin angular momentum operators.
This Hamiltonian is a straightforward extension of the expression \eqref{Hamiltonian}.
Now as an example, we choose $\delta=0$ and $\ket{\psi_0}=[1, 0, 1]^\top/\sqrt{2}$ at $(\theta, \varphi)=(\pi, \pi/2)$.
 In this setting, we obtain the lower bound as
\begin{align}
\label{qutritT}
T \geq T_*=\frac{ \sqrt{1-\cos\Theta_T}}{\Omega+u_{\rm max}},
\end{align}
with
\begin{align}
\label{qutritF}
\cos\Theta_T =\frac{1}{4}\left( \cos\alpha\cos\beta+ \cos\alpha\sin\beta+\cos\beta-\sin\beta  \right)^2.
\end{align}
Thus, from Eq. \eqref{qutritF}, $T_*$ is maximized when $(\alpha, \beta)$ satisfies $\cos\alpha=(\sin\beta-\cos\beta)/(\sin\beta+\cos\beta)$ [Fig. 2 (b)];
here we typically take the following target gates
\begin{align*}
&G(\pi, 0) = \left[
    \begin{array}{ccc}
     -1   & 0 & 0 \\
     0&  -1  &  0 \\
      0 &  0 &  1
    \end{array}
  \right], \ 
G(0, \pi/2) = \left[
    \begin{array}{ccc}
   0 &  0& 1  \\
    0 &   1 &  0 \\
   -1 & 0  &  0
    \end{array}
  \right], \\   
&  G(\pi, \pi) = \left[
    \begin{array}{ccc}
    1 & 0 &0  \\
    0&   -1 & 0  \\
    0&  0 &  -1
    \end{array}
  \right].
\end{align*}
As noted in the previous subsection, the final states generated by the above gates are orthogonal to $\ket{\psi_0}$, 
and thus implementing such goal gates takes longest time.
Moreover, let us particularly consider the spin-up transformation $\ket{\psi_0}\to \ket{E}$, which is realized by $G(0, \pi/4)$ or $G(\pi, 3\pi/4)$.
From Fig. 2 (b), it is immediately found that this transformation takes $T=0.5$ at least.
In this way, we can roughly estimate the time to implement the target gate without solving any equations.

\begin{figure}[tb]
\centering
\includegraphics[width=8.5cm]{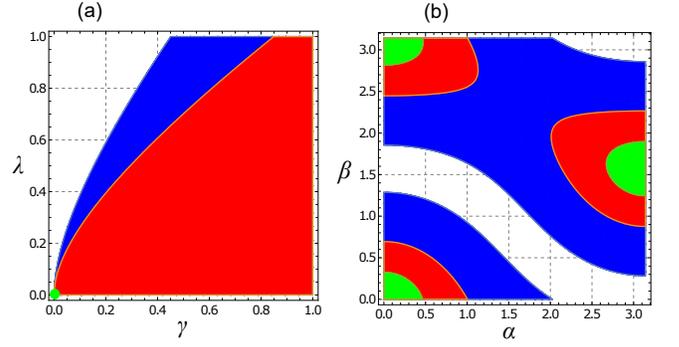}
\caption{(a) Reachable sets of $\lambda$ as a function of $\gamma$ in $T=0.5$ 
when the initial state is $\ket{\Psi^+}$ (blue area), $\ket{\Phi^\pm}$ (red area), and $\ket{\Phi^-}$ (green dot at the origin).
(b) Reachable set of the target gate $G(\alpha, \beta)$ when $\theta=\pi/4$ in unit of $\Omega=u_{\rm max}=1$
for $T=0.3$ (green area), $T=0.5$ (red area), and $T=0.8$ (blue area).
}
\end{figure}

\section{Conclusion}
Motivated by the work by Arenz et al., in this paper we have characterized the reachable set of Markovian open quantum systems under a given control setting based on the QSL.
In order to make this analysis tractable and rigorous, it is important that the QSL is explicitly computable and tighter than other bounds;
the QSL used in this paper indeed satisfies these conditions. 
Thanks to this features, as shown in Sec. III, we can clarify the set of all final states in a given control time 
and roughly estimate the time for the controlled state existng in a certain region from the initial state without solving any equations.
An important remaining work is to extend this approach to more strategic control cases, 
e.g., the measurement-based feedback control.
\\
\ \ \ This work was supported by MEXT Quantum Leap Flagship Program Grant 
Number JPMXS0118067285 and JPMXS0120319794.

\appendix

\section{Proof of QSL}

We begin with taking the time derivative of $\Theta_{t}$:

\begin{align}
\label{theorem proof eq 1}
\frac{d\Theta_{t}}{dt}&=\frac{-1}{\sqrt{1-{\rm Tr}(\rho_{0}\rho_{t})^{2}}} 
\cdot {\rm Tr}\left( \rho_{0} \frac{d\rho_{t}}{dt} \right) \notag \\
&=\frac{1}{ \sin\Theta_{t} }  {\rm Tr}\left\{ \left( i[\rho_{0}, H]- \mathcal{D}^{\dagger}[M]\rho_0 \right)\rho_t \right\},
\end{align}

To have an upper bound of the rightmost side of \eqref{theorem proof eq 1}, 
we often use the Schwarz inequality for matrices $X$ and $Y$: 
\begin{equation}
\label{Schwarz} 
     \left|{\rm Tr}(X^{\dagger}Y)\right| \leq \|X\|_{\rm F}\|Y\|_{\rm F}.
\end{equation}
Also the following inequality is often used:
\begin{align*}
\|\rho_{t}-\rho_{0}\|_{\rm F} &=\sqrt{{\rm Tr}\big[\left(\rho_{t} -\rho_{0}\right)^{2}\big]} =\sqrt{{\rm Tr}\left( \rho_{t}^2-2\rho_{t} \rho_{0}+\rho_{0}^{2}\right)}  \\
         & \leq \sqrt{2-2{\rm Tr}(\rho_{t}\rho_{0})} =\sqrt{2-2\cos\Theta_t},
\end{align*}

where ${\rm Tr}(\rho_{t}^{2})\leq 1$ and ${\rm Tr}(\rho_{0}^{2})=1$ are used.
Using these inequalities, the rightmost side of \eqref{theorem proof eq 1} is upper 
bounded by 
\begin{align}
\label{theorem proof eq 2}
& {\rm Tr}\{ \left(i[\rho_{0}, H] -  \mathcal{D}^{\dagger}[M]\rho_0 \right) \rho_{t} \}  \notag \\
& ={\rm Tr}\{ \left(i[\rho_{0}, H]- \mathcal{D}^{\dagger}[M]\rho_0 \right)(\rho_{t}-\rho_{0})\}
 - {\rm Tr}(\rho_{0} \mathcal{D}^{\dagger}[M]\rho_0 ) \notag\\
&={\rm Tr}\{ \left(i[H, \rho_{0}]+ \mathcal{D}^{\dagger}[M]\rho_0 \right)(\rho_{0}-\rho_{t})\} 
+ {\rm Tr}(M^{\dagger}M\rho_{0}) \notag\\
&\ \ \ - {\rm Tr}(M^{\dagger}\rho_{0}M\rho_{0})    \notag\\
& \leq \|i[H, \rho_{0}]  + \mathcal{D}^{\dagger}[M]\rho_0 \|_{\rm F} \cdot 
\|\rho_{t}-\rho_{0}\|_{\rm F} + \|M\ket{\psi_{0}}\|^{2}       \notag \\
&\ \ \ -|\bra{\psi_{0}}M\ket{\psi_{0}}|^{2}  \notag \\
&\leq \sqrt{2}\| i[H, \rho_{0}] + \mathcal{D}^{\dagger}[M]\rho_0 \|_{\rm F}
\sqrt{1-\cos\Theta_t}+\|M\ket{\psi_{0}}\|^{2}   \notag\\
& \ \ \    -|\bra{\psi_{0}}M\ket{\psi_{0}}|^{2}  \notag\\
&\leq \sqrt{2}\left(  \| i[H, \rho_0]+\mathcal{D}^{\dagger}[M]\rho_0 \|_{\rm F}\right)\sqrt{1-\cos\Theta_t}   \notag\\
& \ \ \ + \|M\ket{\psi_{0}}\|^{2} -|\bra{\psi_{0}}M\ket{\psi_{0}}|^{2}.
\end{align}
From Eqs. \eqref{theorem proof eq 1} and \eqref{theorem proof eq 2}, we have 

\begin{eqnarray}
\label{theorem proof eq 3}
\frac{d\Theta_{t}}{dt} \leq \frac{1}{\sin\Theta_{t}  }
               \left( \mathcal{A}\sqrt{1-\cos\Theta_t}  +\mathcal{E} \right),
\end{eqnarray}
where
\begin{eqnarray*}
     \mathcal{A}&=&\sqrt{2} \left( \|i[H, \rho_0]+ \mathcal{D}^{\dagger}[M]\rho_0 \|_{\rm F}\right),  \\
     \mathcal{E}&=&\|M\ket{\psi_{0}}\|^{2}-|\bra{\psi_{0}}M\ket{\psi_{0}}|^{2}.
\end{eqnarray*}
Finally, by integrating \eqref{theorem proof eq 3} from $t=0$ to $T$, 
we end up with the  lower bound \eqref{QSL}.

\end{document}